\documentclass[10pt]{JHEP3}

\usepackage{amssymb,amsmath}
\usepackage{graphics}
\usepackage{epsfig}
\usepackage{amsfonts}

\usepackage{fancyhdr}

\usepackage{amscd}

\def\be{\begin{eqnarray}}
\def\ee{\end{eqnarray}}

\def\g{\gamma}
\def\e{\epsilon}

\def\half{\frac{1}{2}}
\def\d{\partial}
\def\a{\alpha}
\def\b{\beta}

\newcommand{\idn}{{1\relax{\kern-.35em}1}}

\def\a{\alpha}

\def\b{\beta}
\def\g{\gamma}
\def\e{\epsilon}

\title{On fluctuations of closed string tachyon solitons}
\author{
Shlomo S. Razamat\\
Department of Physics,\\
Technion, Israel Institute of Technology,\\
Haifa 32000, Israel\\
E-mail:\email{razamat@physics.technion.ac.il}
}

\abstract{We discuss fluctuations on solitons in the dilaton/graviton/tachyon system using the low energy effective
field theory approach. It is shown that closed string solitons are free of tachyons in this approximation, regardless of
the exact shape of the tachyon potential.}

\keywords{Closed string tachyons}

\begin{document}

\section{Introduction}

One of the most interesting features of string theory is the appearance, in some cases, of tachyons.
Although much progress has been made
in understanding the dynamics of open string tachyons 
\cite{Sen:2004nf, Schnabl},
and (semi)localized closed string tachyons (for an incomplete list of references see 
\cite{Adams:2001sv, Headrick:2004hz,Vafa:2001ra,Okawa:2004rh,Adams:2005rb,
Horowitz:2005vp,Silverstein1,Bergman:2005qf,Berkooz1,Ross1,Berkooz2}),
bulk closed string tachyons have remained a mystery.

Building on the success in the open string and localized closed string cases one can formulate 
a natural two-part conjecture for bulk closed string tachyons: 1. 
 The ground state contains no degrees of freedom, and 2.
 solitons correspond to lower-dimensional sub-critical closed string backgrounds.\footnote{Indirect evidence
for this conjecture was given in the context of p-Adic strings by Moeller and Schnabl \cite{Moeller:2003gg}. Consult \cite{Zwiebach1}
for a string field theory approach for the problem, and \cite{Yang2} for time dependent
solutions.
See also \cite{hellerman} for a discussion of bulk closed string tachyon condensation
in super-critical string.}
 
In \cite{Bergman:2006pd} we investigated the second part of the conjecture for the bulk closed string
tachyon of the bosonic string. 
We claimed that a co-dimension one
soliton solution of the 26-dimensional closed bosonic string theory describes
the flat linear dilaton background of the 25-dimensional sub-critical bosonic string theory.
 To obtain this we  formulated the problem in the
 low-energy effective theory of 
the tachyon, dilaton and graviton.\footnote{See \cite{Suyama} for similar calculations including the $B$ field}
To specify the action we have to know the form of the tachyon potential. 
However some general features of the soliton solution are independent of
the specific potential. 

The low energy effective theory of the gravity-dilaton-tachyon system
in the closed bosonic string (or the NSNS sector of the Type 0 superstring)
is given in the string frame by\footnote{Note that the metric notations in this note are $(+,-,...,-)$ and
differ from \cite{Bergman:2006pd}.}
\be\label{action}
S=\frac{1}{2\kappa^2}\int d^{D}x\sqrt{-g}\,
 e^{-2\Phi}\left(-R-4(\partial_\mu\Phi)^2+(\partial_\mu T)^2-2V(T)\right)\, ,
\ee
where $V(T)=\half m^2 T^2 + \cdots$. We can look for co-dimension one solitonic solutions
supported by this action.  These solutions are specified by a static tachyon profile $\bar{T}(x_{D-1})$,
such that $\bar{T}(0)=0$ and $\bar{T}(x_{D-1})$ approaches a minimum of $V(T)$ as 
$x_1\rightarrow \pm\infty$. If there is a unique minimum, the soliton is a lump,
and if there are degenerate minima the soliton is a kink. Further, assuming a simple form for the metric ($\mu,\nu = 0,1,\ldots,D-2$),
\be\label{metric}
ds^2=-dx_{D-1}^2+a(x_{D-1})^2\eta_{\mu\nu}dx^\mu dx^\nu \,,
\ee
one can show that the metric in the string frame should be flat, i.e. $a'=0$, and that
the dilaton profile has to have the following form ($x_{D-2}$ is a direction transverse to the soliton) 
\be
\label{total_dilaton}
 \bar{\Phi}(x_{D-1},x_{D-2}) = D(x_{D-1}) + qx_{D-2} \qquad ,
\ee for some $q$.\footnote{Essentially, the equations of motion require that the product $q a' $  will vanish. However,
as we want to interpret the soliton as the sub-critical string we take the solution with non-vanishing linear dilaton
profile.} The dilaton must therefore be linear in the soliton "world-volume" coordinates.
In this case the equations of motion reduce to
\be
 2 D'' - (\bar{T}')^2 &=& 0 \label{gravity} \\
\bar{T}'' - 2 D' \bar{T}'-V'(\bar{T}) &=& 0 \label{tachyon}\\
D'' - 2(D')^2 - 2q^2-V(\bar{T}) &=& 0  \,\label{dilaton}.
\ee
This is an over-determined set of equations for the two fields and the constant $q$. Consistency of
these equations and the assumption that there is only one solitonic solution (or a discrete set of solutions) supported
by this system fixes the value of the constant $q$ in terms of the parameters of the tachyon potential.

Moreover, one can show that the qualitative behavior of the dilaton profile in the direction of the soliton, $x_{D-1}$, is generic
and does not depend on the details of the tachyon potential.  Interpreting the tachyon equation of motion above as an
equation for a point particle in inverted potential, $-V$, with a friction term given by $-2D'$, and noting that the second
derivative of the dilaton, $D''$, should be non-negative, one concludes that the dilaton has to grow as $x_{D-1}\rightarrow \pm\infty$.
This implies that the Einstein frame metric vanishes away from the core of the soliton, and the space-time effectively localizes on the $(D-1)$-dimensional 
worldvolume of the soliton.
The whole picture is consistent with the identification of the soliton
as the flat linear-dilaton background of the $(D-1)$-dimensional string theory.

In this note we will analyze fluctuations around these solitons. For the soliton to describe a lower 
dimensional string theory one expects that the fluctuation spectrum will still contain a tachyon, since a lower dimensional
string theory is still unstable. However, the mass squared of the tachyon living on the soliton should decrease in absolute value
as the number of dimensions decreases. In what follows we will find that the mass of the would-be tachyon living on the soliton
is essentially non tachyonic in the low energy effective field theory approximation.

This note is organized as follows.
In section \ref{fluct_sect} we will discuss fluctuations around solitons for an unspecified,  generic class of tachyon potentials.
In section \ref{general_sect} we will discuss generic features of the fluctuations and we will illustrate these features on a simple model of quadratic tachyon potential in section
\ref{quadratic_sect}. In section \ref{disc_sect} we will summarize and discuss the results. 
Finally, in appendix \ref{app_quad} we will discuss the quadratic tachyon
potential neglecting the gravity fluctuations.

\section{Fluctuations on the solitons}\label{fluct_sect}
We will analyze the system of a graviton, dilaton and a scalar field which we will refer to as a tachyon although
it does not have to posses a negative mass squared.\footnote{Consult \cite{Karch} for similar calculations.}
The action in the Einstein frame is
\be
S&=&\frac{1}{2\kappa^2}\int d^Dx\sqrt{-g}\biggl[-R+M^{IJ} \partial_\mu\Phi_I\partial^\mu\Phi_J-2\tilde V(\Phi_I)\biggr].
\ee We have defined
\be
&&M_{11}=\frac{4}{D-2},\qquad M_{22}=1,\qquad M_{12}=M_{21}=0,\qquad \tilde V(\Phi_I)=e^{\frac{4}{D-2}\Phi}V(T).
\ee
 $\Phi_1=\phi_1+\varphi_1$ is the dilaton and
$\Phi_2=\phi_2+\varphi_2$ is the tachyon. The $\phi_i$ are  the background configurations
of the fields and $\varphi_i$ are the fluctuations of the fields around that background.
We will denote $$x_{D-1}=x,\qquad x_{D-2}=y.$$
The prime denotes a derivative with respect to $x$, and the dot denotes a derivative with respect to $y$. 
In all the models we deal with \cite{Bergman:2006pd}$$\phi_2=\phi_2(x),\qquad \frac{2}{D-2}\phi_1=A(x)+B(y),\qquad B(y)=\frac{2q}{D-2}\cdot y,$$ 
i.e. the dilaton background is linear in one of the directions, $y$, and has some profile in direction $x$. The tachyon background depends only on the coordinate $x$.
General metric fluctuations in our background \eqref{metric} are given by \be ds^2=e^{-2S}(\eta_{\mu\nu}+2h_{\mu\nu})dx^\mu dx^\nu,\ee
where $S=A+B$. Define the fluctuations of the Christoffel symbols
\be
\Gamma^{\rho}_{\nu \mu}&=&\Gamma^{(0)\rho}_{\nu \mu}+C^{\rho}_{\nu \mu}.
\ee
These are calculated with an ansatz above to give
\be
\Gamma^{(0)\rho}_{\nu \mu}&=&-\d_\mu{S}\delta^{\rho}_\nu-\d_\nu{S}\delta^{\rho}_\mu
+\d^\rho S\eta_{\mu\nu},\\
C^{\rho}_{\nu \mu}&=&\eta^{\rho\a}(\d_\mu h_{\nu\a}+\d_\nu h_{\mu\a}
-\d_\a h_{\mu\nu})+2\d^\rho Sh_{\mu\nu}-2\d_\a S h^{\rho\a}\eta_{\mu\nu}\nonumber
\ee
In what follows we raise and lower indices with the flat metric. The fluctuations of the Ricci tensor and the stress tensors are given by
\be
\delta R_{\mu\nu}&=&
D^\a (\d_\mu h_{\nu\a}+\d_\nu h_{\mu\a})-D^\a\d_\a h_{\mu\nu}  -\d_\mu\d_\nu h+\d^\rho S\d_\rho h \eta_{\mu\nu}-
2D_\rho(\d_\a S h^{\rho\a})\eta_{\mu\nu}
-\frac{4}{D-2}V h_{\mu\nu}\nonumber\\
\delta\tilde T_{\mu\nu}&=&M_{IJ}\d_\mu\phi_I\d_\nu\varphi_J+M_{IJ}\d_\nu\phi_I\d_\mu\varphi_J-\frac{2}{D-2}\eta_{\mu\nu}V_I\varphi_I-\frac{4}{D-2}V h_{\mu\nu}
\ee where we have defined
$$\tilde T_{\mu\nu}=T_{\mu\nu}-\frac{1}{D-2}g_{\mu\nu}T,\qquad D_\a=\d_\a-(D-2)\d_\a S.$$ To solve for the spectrum of fluctuations on the soliton 
we have to solve the Einstein equations of motion, $$\delta R_{\mu\nu}=  \delta\tilde T_{\mu\nu},$$ as well
as the equations of motion for the tachyon and the dilaton. We
partially fix the gauge by setting (index $i$ runs over $0,1,...,D-3$)
\be\label{gauge} h_{ix}=0,\qquad h_{xy}=0,\qquad \d^i h_{yi}=0,\qquad \biggl[\;\square = \d^i\d_i,\quad \tilde h\equiv h^i_{\;i}\;\biggr].\ee The equation of motion for the fluctuations of the
tachyon and the dilaton in this gauge are given by 
\be\label{field1}
 &&M_{IJ}\biggl\{ D^\rho \d_\rho \varphi_J -2h_{xx}D_x\d_x\phi_J-2h_{yy}D_y\d_y\phi_J
 -2\left[(\half h'+h_{xx}')\phi_J'+(\half \dot h+\dot h_{yy})\dot \phi_J\right]\biggr\}=-V_{IJ}\varphi_J.\nonumber
 \ee
 The quantities appearing in the above equation are given by
 \be
 &&V_1=\frac{4}{D-2}V,\qquad V_2= V',\qquad V_{11}=\biggl(\frac{4}{D-2}\biggr)^2V,\\
 &&V_{22}=V'',\qquad V_{12}=V_{21}=\frac{4}{D-2}V'.\nonumber
 \ee
The background field equations (\ref{tachyon},\,\ref{dilaton}), can be written as 
 \be\label{field0}
  V_I -M_{IJ}(\phi_J''+\ddot\phi_J)+(D-2)M_{IJ}(A'\phi_J'+\dot B\dot\phi_J)=0.
 \ee We define the following useful quantities
 \be\label{Rdef}
 R_y&\equiv&D^yh_{yy}-\dot h
- M_{IJ}\dot\phi_I\varphi_J+\frac{1}{D-2}\dot{\tilde h}\,\, ,\\
R_x&\equiv&D^x h_{xx} -h'
- M_{IJ}\phi'_I\varphi_J+\frac{1}{D-2}\tilde h'\,\, .\nonumber
\ee 
Using the definitions above the following components of the Einstein equations are given by
\be\label{hprop}
\{x,y\}\qquad &:& \qquad\;\;-\left( \dot R_x+R'_y\right)=\d_x\d_y\biggl[h-\frac{2}{D-2}\tilde h\biggr],\\
\{x,i\}\qquad &:& \qquad\;\;\d_i R_x=-D^\a h'_{i\a}+\frac{1}{D-2} \d_i\tilde h',\nonumber\\
\{y,i\}\qquad &:& \qquad\;\;\d_i R_y=-D^\a \dot h_{i\a}+\frac{1}{D-2} \d_i\dot{\tilde h}+D^\a\d_\a h_{iy}.\nonumber
\ee
One can verify that the two scalar equations \eqref{field1}  imply that $R_x=R_y=0$. 
Thus, the $\{x,i\}$ equation becomes
\be\label{condEq}
-D^\a h'_{i\a}+\frac{1}{D-2}\d_i\tilde h'=0.
\ee
We can use part of the residual gauge symmetry to set also 
\be\label{condEq2}
-D^\a  h_{i\a}+\frac{1}{D-2}\d_i{\tilde h}=0.
\ee
Plugging this into the $\{i,y\}$ equation we obtain
\be\label{vector}
D^\rho\d_\rho h_{iy}=0.
\ee Note that this is a simple laplacian of a scalar in our curved background.
Using the above remaining components of the Einstein equations are
\be\label{GHF1}
\{x,x\}&:& D^x h'_{xx}-\half\left[ D^\rho\d_\rho h_{xx}+h''+\d^\rho S\d_\rho h\right]
- M_{IJ}\phi_I'\varphi_J'-\frac{V_I\varphi_I }{D-2}
+D_x (A'h_{xx})+D_y (\dot B h_{y y})=0,\nonumber\\
\{y,y\}&:& D^y \dot h_{yy}-\half\left[D^\rho\d_\rho h_{yy}+\ddot h+\d^\rho S\d_\rho h\right]
- M_{IJ}\dot \phi_I\dot \varphi_J-\frac{ V_I\varphi_I}{D-2}
+D_x (A'h_{xx})+D_y (\dot B h_{y y})=0,\nonumber\\
\{i,j\}&:&  -D^\rho\d_\rho h_{ij}
+ \biggl[\frac{2V_I\varphi_I}{D-2}+\d^\rho S\d_\rho h-2D_x (A'h_{xx})-2D_y (\dot B h_{y y})\biggr]\eta_{ij}=0.
\ee  Projecting on the  traceless ( $\tilde h_{ij}=h_{ij}-\frac{1}{D-2}\,\tilde h\, \eta_{ij}$ )
part we get again
\be\label{graviton}
D^\rho\d_\rho \tilde h_{ij}=0.
\ee
This equation along with \eqref{vector} describe the tensor and the vector excitations on the soliton.
 For instance, the lowest component of  \eqref{graviton} describes a $25$-dimensional graviton confined to the soliton world-volume.
 
Further, using \eqref{GHF1} one can compute the following 
\be
\frac{D-4}{D-2}\{i,i\}-\{x,x\}-\{y,y\}=0,\qquad\to\qquad  \square\biggl[ h-\frac{2}{D-2} \tilde h\biggr]=0,
\ee in agreement with the first equation in \eqref{hprop}. We can use the remaining residual gauge symmetry
to set $h-\frac{2}{D-2} \tilde h=0$. 
Note that the above relation implies that the three scalars,
$h_{xx}$, $h_{yy}$ and $h$, are linearly dependent.
To find the fluctuations of the scalar modes we define
\be
Q=(D-2)h_{yy}+\tilde h,\qquad P=(D-2)h_{xx}+\tilde h.
\ee
From here using \eqref{Rdef} and (\ref{GHF1}) we get simple equations for $P$ and $Q$
\be
D^\rho\d_\rho Q&=&0,\label{finalQ}\\
 D^\rho\d_\rho P &=&
-2\frac{\phi_2''}{\phi_2'}P'-2\frac{\phi_1''\phi_2'-\phi_2''\phi_1'}{\phi_2'}\biggl[2P+\frac{1}{q}\dot{Q}-2Q\biggr].\label{finP}
\ee
In what follows we will be interested in the scalar spectrum on the soliton. To summarize, the tensor and the vector
fluctuations on the soliton are given in terms of a simple laplace equation in our curved background, equations (\ref{vector}) and (\ref{graviton}).
 The spectrum of scalar
fluctuations consists of two independent fields,
 which we can choose to be $P$ and $Q$. Thus to
obtain the scalar spectrum of the soliton we have to solve (\ref{finalQ}) and (\ref{finP}).

\section{General features of the spectrum}\label{general_sect}

In this section we will derive some general features of the scalar spectrum of fluctuations. 
 First, let us denote
 \be
 \hat Q &=& \biggl[\frac{1}{q}\d_y-1\biggr] e^{-\phi_1-q\, y}Q,\qquad \bar P=\frac{1}{\phi'_2}e^{-\phi_1-q\, y}P,\\
 H_P&=&-\d_x^2+\biggl\{\left(\frac{\phi_2''}{\phi_2'}-\phi_1'\right)^2-\left(\frac{\phi_2''}{\phi_2'}-\phi_1'\right)'+
 2\phi_1''+q^2\biggr\},\nonumber\\
H_Q&=&-\d_x^2+\left\{(\phi_1')^2-\phi''_1+q^2\right\},\qquad F= 2\left(\frac{\phi_1'}{\phi_2'}\right)'.\nonumber
 \ee
 Using these definitions the equations (\ref{finalQ}) and (\ref{finP}) take the following form
\be
\square \bar P -\ddot{\bar P}&=& -H_P \bar P-F \hat Q,\\
\square \hat Q -\ddot{\hat Q}&=& -H_Q \hat Q.\nonumber 
\ee
 To diagonalize these we write 
\be
\hat P=\bar P+\hat L \hat Q, 
\ee where $\hat L$ is a linear operator that we demand to satisfy
 \be\label{diag_eq}
 F=H_P\hat L-\hat L H_Q.
 \ee In general this equation for $\hat L$ is hard to solve, but we will see an example
 for which the equation \eqref{diag_eq} simplifies significantly. Finally, the diagonalized fluctuation equations take the
 following form
 \be\label{PQdiag}
 \square \hat P -\ddot{\hat P}=-H_P\hat P,\qquad
 \square \hat Q -\ddot{\hat Q}=-H_Q\hat Q.
 \ee The mass spectrum is given by the eigenvalues of the Schrodinger-like operators $H_{P,Q}$.
  Note that the $Q$ equation is exactly the dilaton fluctuation equation if we neglect mixing completely \cite{Bergman:2006pd}.
  However, here $Q$ is a combination of the dilaton and the gravity fields.
 
\medskip
 \noindent Some general features of the fluctuation spectrum are easy to extract.
One can immediately conclude from above that there is no tachyon on the soliton in our effective field theory for any kind
 of tachyon potential. Note that \eqref{PQdiag} can be written as 
 \be\label{PQAAD}
 &&\square \hat P -\ddot{\hat P}=-\left[a_P^\dagger a_P+
 (\phi_2')^2+q^2\right]\hat P,\\
 &&\square \hat Q -\ddot{\hat Q}=-\left[a_Q^\dagger a_Q+q^2\right]\,\,\hat Q,\nonumber
 \ee where we have defined
 \be
 a_P=\d+\frac{\phi_2''}{\phi_2'}-\phi_1',\quad a_P^\dagger=-\d+\frac{\phi_2''}{\phi_2'}-\phi_1',\quad  a_Q=\d+\phi'_1,\quad   a_Q^\dagger=-\d+\phi'_1.
 \ee Note that this immediately implies that the expectation value of the hamiltonian $H_{P,Q}$ on any state is positive definite
 and thus there are no tachyons in the spectrum.
 
Another general feature  is that if $\phi_1$ becomes large at $x\to\pm\infty$ we can obtain the ground state of the $\hat Q$
states. It is simply given by $a_Q\hat Q=0$ condition, giving $\hat Q=e^{-\phi_1}$. If the dilaton does not go to strong coupling at infinities
one can not say anything about the ground state. However, as was noted in \cite{Bergman:2006pd} the dilaton diverges at $x\to\pm\infty$, and thus,
this is always the $\hat Q$ ground state in the cases of the solitons.

\section{Quadratic potential}\label{quadratic_sect}

In this section we will discuss a simple example for which the diagonalization problem is
exactly solvable and one can rigorously obtain the spectrum.
We begin by setting 
\be
\phi''_2=0.
\ee Without loss of generality we set \be \phi_2(x)=2x.\ee Then the diagonalization 
\eqref{diag_eq} is very simple
 \be\hat P =\bar P+\frac{1}{4}\hat Q\ee   Using the dilaton and
 tachyon equations of motion we find that the tachyon potential and the dilaton profile are
 \be
 V(T)=2-2q^2-2\, T^2,\qquad \phi_1(x,y)=x^2+qy.
 \ee
 Note that the physical demand that the potential will vanish for $T=0$ implies $q^2=1$, but we won't restrict to this in what follows. 
 The fluctuation equations are diagonalized
 \be\label{Quad_diag}
 &&\square \hat P -\ddot{\hat P}=-\left[- \hat
 P''+\biggl\{4x^2+6+q^2\biggr\}\hat P\right],\\
 &&\square \bar Q -\ddot{\bar Q}=-\left[- \bar Q''+\biggl\{4x^2-2+q^2\biggr\}\bar Q\right].\nonumber
 \ee
The  RHS are simple harmonic oscillator Hamiltonians which give evenly spaced towers of states with masses squared given by
\be
m_Q^2=4n+q^2,\qquad m_P^2=4n+8+q^2.
\ee
 Thus we have two massive fields with mass squared $q^2,q^2+4$ and doubly degenerate massive spectrum $4n+q^2+8$.
 There is no tachyon.  

\begin{figure}[htbp]
\begin{center}
$\begin{array}{c@{\hspace{0.0in}}c}
\epsfig{file=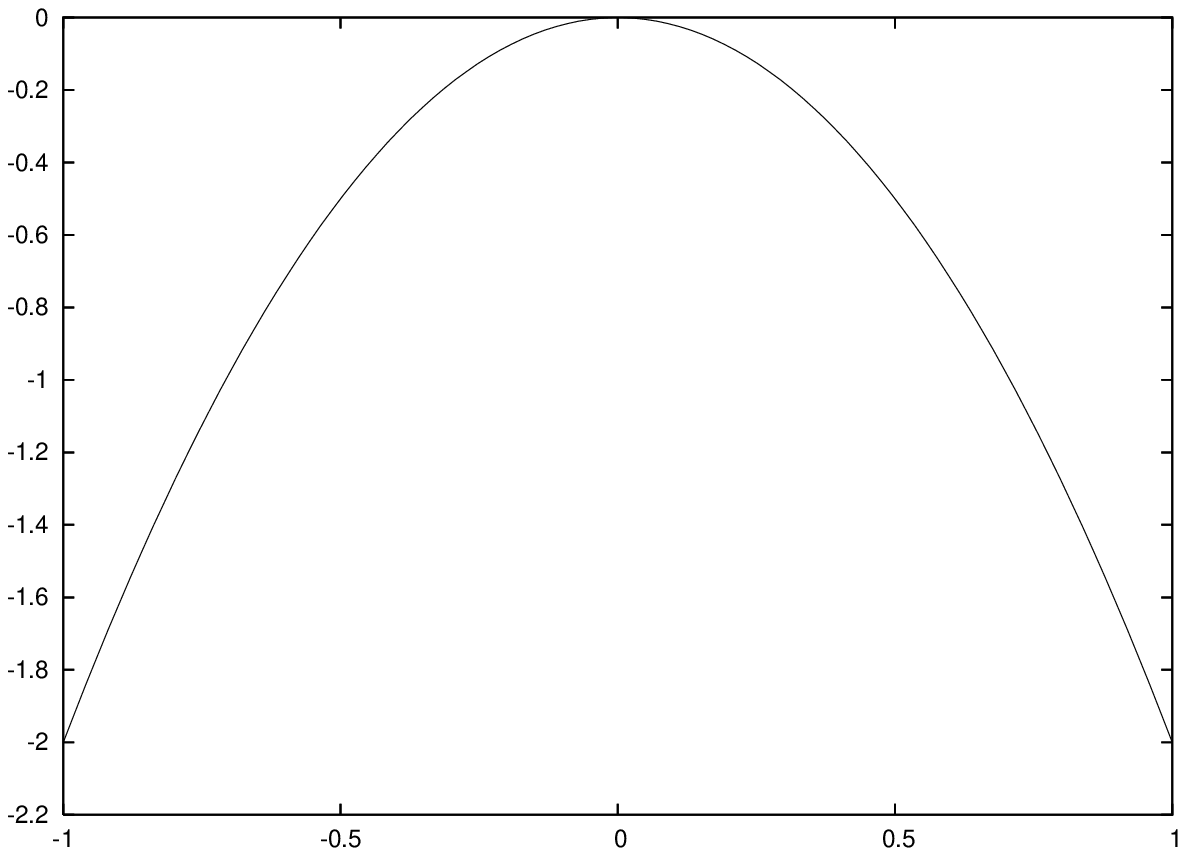, scale=0.57} &
    \epsfig{file=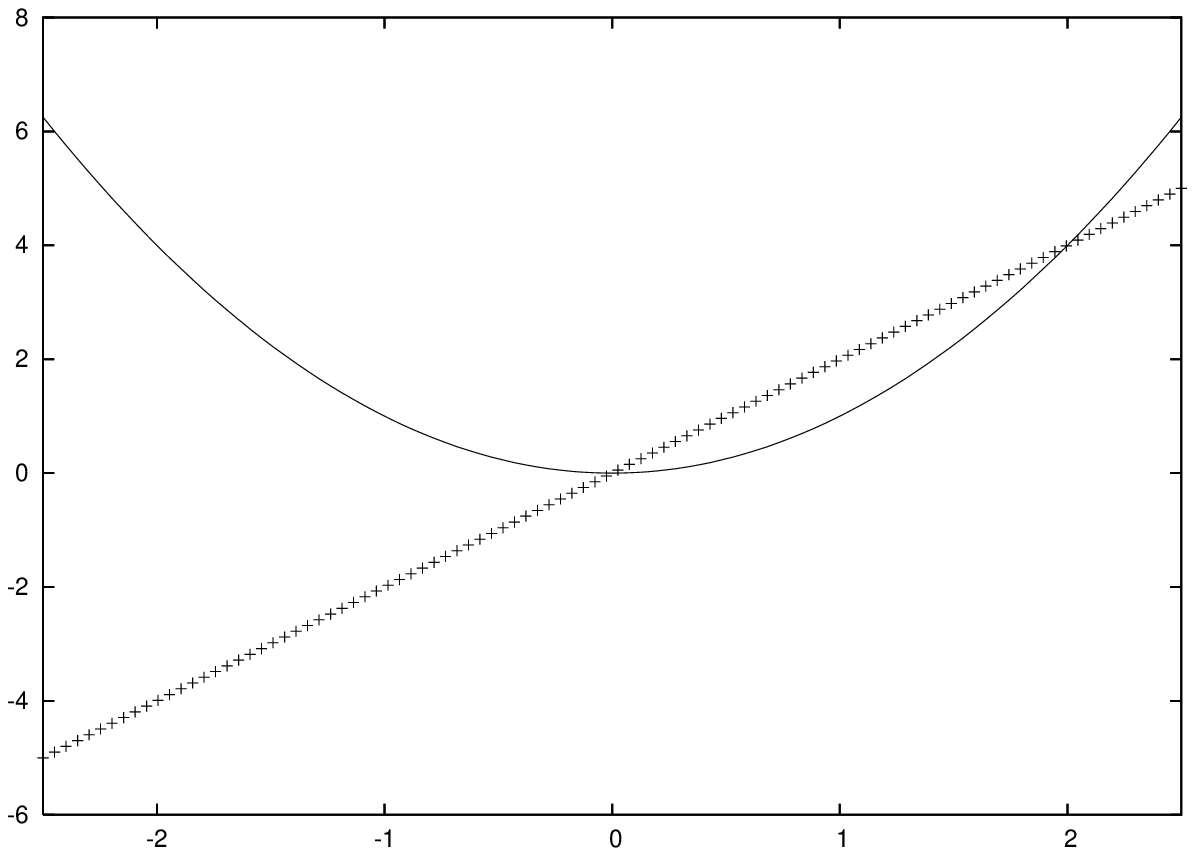, scale=0.57} \\ [0.2cm]
\end{array}$
 \caption{An example of quadratic tachyon potential, $V(T)$, with $q=1$ and the dilaton/tachyon profiles (the dotted line is the tachyon profile, $\phi_2(x)$,
 and the solid line is the dilaton profile , $\phi_1(x)$).
Note that the tachyon is $+\infty$ at $x=+\infty$,
and goes to $-\infty$ at $x=-\infty$, extrapolating between the two vacua. }\label{QuadFig}
\end{center}
\end{figure}

\section{Summary and Discussion}\label{disc_sect}

In this note we have investigated the spectrum of fluctuations on closed string tachyon
solitons. Our motivation for investigating these objects is to interpret them as sub-critical strings. We have 
analyzed the low energy effective action describing the tachyon, dilaton and graviton. 
The main conclusion from the calculations is that there is no tachyonic instability   on
the solitons in our low energy models. There are two possible explanations of this. First,
the tachyon might be restored when we take into account the massive fields in the string spectrum.
If the mixing between the massive and the massless fields (or the tachyon) is strong enough,
a negative mass squared fluctuation might re-appear. Another possible explanation comes from
the observation that our final expressions \eqref{PQdiag} describing the spectrum of
the fluctuations do not depend on the number of space-time dimensions. 
Our analysis did not assume a specific number of dimensions and thus in particular it should be
suitable  for describing a co-dimension one soliton in three dimensions. Following
our general expectations, we interpret this soliton as a $2d$ string. The tachyon of the $2d$ string
 is massless and this is consistent with our findings. Thus, the low energy analysis might be 
suitable only for this case for some reason. It would be very interesting to investigate
 all these issues further.

\section*{Acknowledgments}
 I would like to thank Oren Bergman, Zohar Komargodski and Amos Yarom for useful discussions.
 I would especially like to thank  Andreas Karch for a very useful correspondence.
     This work  is supported in part by the Israel Science Foundation under
 grant no. 568/05.

\appendix

\section{The naive fluctuation spectrum for a quadratic tachyon potential.}\label{app_quad}
In this appendix we will discuss the spectrum on the soliton of a quadratic tachyon potential without taking into account gravity
 fluctuations.
Note that when we neglect gravity we first should specify in what frame we are working: the string frame or the Einstein frame.
The fluctuation spectrum is different in the two frames if we neglect gravity, since changing frames involves non trivial
field redefinitions involving metric. In what follows we will compute the spectrum in the string frame, where we can do it explicitly. 
In section \ref{quadratic_sect} we defined the model:
\be
\phi_1=x^2,\qquad \phi_2=2x,\qquad V(T)=-2T^2, \qquad q^2=1
\ee
In the string frame the fluctuations are given by (see \cite{Bergman:2006pd} for details):
\be
\label{fluctuation_L}
L =
 \left[
 \begin{array}{cc}
 -\partial^2 + \Delta &  - {\phi_2}'(\phi'_1 - \partial) \\
 -  {\phi_2}'(\phi'_1 + \partial) \;\;\; & \partial^2 - \Delta - V''(\bar{T})
 \end{array}
 \right]
\ee
and $\Delta = (\phi'_1)^2 - \phi''_1 + q^2$. Using the above model we obtain:
\be
&&\Delta=4x^2-1,\qquad {\phi_2}'(\phi'_1 + \partial_1)=4(x+\half \partial).
\ee
Define 
\be
H_0=\half\biggl(-\partial^2+4x^2\biggr)=2a^\dagger a+1,\qquad a=x+\half\partial,\qquad a^\dagger=x-\half\partial,\qquad [a,a^\dagger]=1.
\ee We know that the spectrum of this Hamiltonian
is \be E_n=2n+1.\ee
  Further, we note that the mixing term is simply $-8a$. Thus we can write:
\be
L \to
 \left[
 \begin{array}{cc}
 2H_0-1 & -4a^\dagger \\
 - 4a \;\;\; & -(2H_0-5)
 \end{array}
 \right].
\ee
First, neglect the mixing and denote the eigenstates of the dilaton as $\psi^d_n$, each having energy
\be\e^d_n=2E_n-1=4n+1.\ee Denote the tachyon eigenstates as $\psi^t_n$, each having energy
\be\e^t_n=2E_n-5=4n-3.\ee We see that the tachyon has a mass squared $-3$ while the original tachyon had mass squared $-4$.
 Now remember that \be a\psi^{t,d}_n=\sqrt{n}\psi^{t,d}_{n-1},\qquad a^\dagger\psi^{t,d}_n=\sqrt{n+1}\psi^{t,d}_{n+1}\ee
  Next we incorporate the mixing. The lowest state of the dilaton
 $\psi^d_0$ does not mix with anything and has mass squared $\e=1$. Further, $\psi^d_{n+1}$ mixes only with $\psi^t_{n}$,
  and the mixing matrix takes the following form
\be M_n=
 \left[
 \begin{array}{cc}
 4n+5 & -4\sqrt{n+1} \\
 -4\sqrt{n+1} \;\;\; & -4n+3
 \end{array}
 \right]\equiv \left[
 \begin{array}{cc}
 \a_n & \gamma_n \\
 \gamma_n \;\;\; & \b_n
 \end{array}\right]
\ee
Remember that the signs of the tachyon kinetic term and the dilaton kinetic term in the string frame are opposite and thus to find the
 spectrum we have to
 diagonalize the above matrix while keeping the kinetic term unchanged. In this case it means that we have to find a matrix $U$ such that
 \be
 U^\dagger
 \left[
 \begin{array}{cc}
 1 & 0 \\
 0 \;\;\; & -1
 \end{array}\right]U=
\left[
 \begin{array}{cc}
 1 & 0 \\
 0 \;\;\; & -1
 \end{array}\right]\qquad\to\qquad
U=\left[
 \begin{array}{cc}
 \cosh x & \sinh x \\
 \sinh x \;\;\; & \cosh x
 \end{array}\right]
 \ee
We want to solve
\be
M_n=U_n\left[
 \begin{array}{cc}
 a_n & 0 \\
 0 \;\;\; & b_n
 \end{array}\right]U_n
\ee
This implies that:
\be
\g_n=\cosh x_n\sinh x_n(a_n+b_n),\quad \a_n=a_n+\sinh^2 x_n(a_n+b_n), \quad \b_n=b_n+\sinh^2 x_n(a_n+b_n)\nonumber
\ee From here we easily solve
\be
\tanh x_n=\frac{\g_n}{\a_n+\b_n},\qquad a_n-b_n=\a_n-\b_n, \qquad a_n+b_n=\frac{\a_n+\b_n}{\cosh^2 x_n}.\nonumber
\ee
Note that these equations imply that if $\biggl|\frac{\g_n}{\a_n+\b_n}\biggr |\leq 1$ the fields are real.
Otherwise they are imaginary. We can make them real by multiplying  with $i$. This will
flip the sign of the kinetic terms of both fields. From here
 the spectrum  is easily computed. It consists of two towers, $m^2_n=1+3n$ and $\tilde m^2_n=-2+5n$. We have a tachyon and thus
 we see that the disappearance of the tachyon is indeed due to the mixing with gravity which we neglected here.

\end{document}